\documentclass[11pt,a4paper]{article}
\usepackage{jcappub}
\title{Axion-photon Propagation in Magnetized Universe}
\author[a]{Chen Wang}
\author[b]{Dong Lai}
\affiliation[a]{National Astronomical Observatories, Chinese
Academy of Sciences, A20 Datun Road, Chaoyang District, Beijing
100012, China} \affiliation[b]{Cornell Center for Astrophysics and
Planetary Science, Department of Astronomy, Cornell University,
Ithaca, NY 14853, USA}

\emailAdd{wangchen@nao.cas.cn}
\emailAdd{dong@astro.cornell.edu}

\abstract{ Oscillations between photons and axion-like particles
(ALP) travelling in intergalactic magnetic fields have been
invoked to explain a number of astrophysical phenomena, or used to
constrain ALP properties using observations. One example is the
anomalous transparency of the universe to TeV gamma rays.  The
intergalactic magnetic field is usually modeled as patches of
coherent domains, each with a uniform magnetic field, but the
field orientation changes randomly from one domain to the next
(``discrete-$\varphi$ model'').  We show in this paper that in
more realistic situations, when the magnetic field direction
varies continuously along the propagation path, the photon-to-ALP
conversion probability $P$ can be significantly different from the
discrete-$\varphi$ model.  In particular, $P$ has a distinct
dependence on the photon energy and ALP mass, and can be as large
as $100\%$.  This result can affect previous constraints on ALP
properties based on ALP-photon propagation in intergalactic
magnetic fields, such as TeV photons from distant Active Galactic
Nucleus.}

\arxivnumber{1511.03380}

\newcommand{\go}{\mathrel{\raise.3ex\hbox{$>$}\mkern-14mu
             \lower0.6ex\hbox{$\sim$}}}
\newcommand{\lo}{\mathrel{\raise.3ex\hbox{$<$}\mkern-14mu
             \lower0.6ex\hbox{$\sim$}}}
\newcommand{\lp}{\left(}
\newcommand{\rp}{\right)}

\newcommand{\vecB}{{\bf B}}

\newcommand{\vecE}{{\bf E}}
\newcommand{\calE}{{\cal E}}
\newcommand{\intd}{{\rm d}}

\begin{document}

\maketitle
\flushbottom

\section{Introduction.}

Axion is particle first introduced to solve the strong CP problem
~\cite{pq77}. Axion-like particles (ALPs) also appear in many
theoretically well-motivated extensions of the standard model of
particle physics~\cite{jae10}. A general property of ALPs
(represented by the field $a$) is that they can couple to photons
(represented by ${\bf E}$) in the presence of an external magnetic
field $\vecB$ through the interaction Lagrangian ${\cal L}=g\,a\,
{\bf E}\cdot{\bf B}$.  While for axions there exists a relation
between the coupling constant $g$ and the axion mass $m_a$, in
general $g$ and $m_a$ are unrelated for ALPs.

As a result of the photon-ALP coupling, a photon can oscillate
into an ALP and vise versa in an external magnetic field. Such
ALP-photon oscillations have been invoked to explain a variety of
astrophysical phenomena, or conversely used to constrain the
properties of ALPs using observations~\cite{mir08}. Examples
include the apparent dimming of distant
supernovae~\cite{csa02,mir05,ost05}, spectral distortions of the
cosmic microwave background~\cite{dia14,csa14}, and the dispersion
of QSO spectra \cite{ost05}, et al. Recently, ALP-photon
oscillation has been used to explain anomalous lack of opacity of
the Universe to gamma rays: high energy gamma ray photons from
Active Galactic Nuclei (AGNs) at cosmological distances have been
detected by HESS, MAGIC and
Fermi~\cite{aha06,maz07,alb08,ack12,abr13}. These photons can
suffer significant attenuation before reaching Earth due to
electron-positron pair production on the extragalactic background
infrared radiation. Several analysis suggest that the Universe
appears more transparent than expected based on recent
extra-galactic background light models(~\cite{dea11,wou12,bru13};
however see \cite{abr13,bit15}). A possible explanation to the
transparency problem is that because of the ALP-photon mixing,
radiation from AGNs travels in the form of ALPs on a significant
fraction of distance (without producing pairs) and converts back
to photons before their
detections~\cite{sim08,bur09,mir09,fai11,hor12,abr13b,bru13,mey14,wou14,gal15}.
Another example concerns the possibilities that the recent
observed 3.55keV photon line~\cite{bul14,boy14} may arise from
dark matter decay to ALPs and then convert to photons due to
oscillations in the magnetic field of M31 and the Milky
Way~\cite{hig14,jae14,cic14,aba14,con14}.

ALP-photon propagation over cosmological distance is strongly affected
by the magnetic field structure. The primordial extragalactic magnetic
field is most likely random and could described as patches of coherent
domains with a typical magnitude upper limit of a few nG~\cite{ade15}
and scale length of order a few Mpc~\cite{gr01}. Previous studies have
adopted a simple model, in which the magnetic field is uniform in each
domain, but the field orientation (characterized by the angle
$\varphi$) changes in a random fashion from one domain to the next.
Note that in this ``discrete-$\varphi$'' model, the photon-to-ALP
conversion probability in each domain $P_{\rm ad}$ can be easily
derived [see Eq.~(\ref{eq:Pad})] (since $\varphi$ is constant in each
domain). Assuming that $\varphi$'s for different domains are random,
Grossman et al.~\cite{gro02} then derived an expression for the
photon-to-ALP conversion probability through a large number of domains
[see Eq.~(\ref{eq:PadN})], and this expression has been widely used in
many previous studies.

In realistic situations, the magnetic field and its orientation angle
$\varphi$ should vary continuously across neighboring domains. In
fact, for a wide range of interesting ALP/magnetic field parameter
space, the variation of $\varphi$ with distance is sufficiently rapid
that it cannot be neglected in almost all regions along the path of
propagation.  We show in this paper that a proper treatment of the
random variation of the intergalactic magnetic field gives a
qualitatively different result for the photon-to-ALP conversion
probability compared to that obtained in the ``discrete-$\varphi$''
model.

For concreteness, we will focus on TeV photon-ALP propagating
through intergalactic medium over cosmological distances, but our
analysis and method can be easily re-scaled to other situations
such as the Milky way or galaxy clusters, as well as for different
photon energies.

\section{Equations}

The evolution equation of the photon electric field $\vecE$ and
ALP field $a$ of a given angular frequency $\omega$ or energy
$\calE$ (so that $\vecE,a\propto e^{i\omega t}$), expressed in a
fixed Cartesian coordinates $xyz$ (with the $z$-axis along the
direction of propagation), takes the form
\begin{equation}
i\lp\begin{array}{c} a'  \\ E_x' \\ E_y'\\
\end{array}\rp = \lp \begin{array}{ccc}
\omega+\Delta_a  & \Delta_M\cos\varphi & \Delta_M\sin\varphi \\
\Delta_M\cos\varphi & \omega +\Delta_{\rm pl} & 0  \\
\Delta_M\sin\varphi & 0  & \omega+\Delta_{\rm pl}  \\ \end{array} \rp
\lp\begin{array}{c} a  \\ E_x \\ E_y\\ \end{array}\rp.
\label{eq:eveq0}
\end{equation}
Here the superscript $'$ stands for $\intd/\intd z$, and $\varphi$ is
the azimuthal angle of the magnetic field $\vecB$ (more precisely,
$\varphi$ is the angle between $\vecB_{\rm tr}$, the projection of
$\vecB$ in the $xy$-plane, and the $x$-axis).  The ALP-mass-related
parameter $\Delta_a$ and the ALP-photon coupling parameter $\Delta_M$
are given by
\begin{eqnarray}
\Delta_a&=&-\frac{m_a^2}{2\omega}=-7.83\times10^{-2}\calE_1^{-1}m_1^2~{\rm Mpc}^{-1},
\label{eq:da}\\
\Delta_M&=&\frac{1}{2}gB_{\rm tr}=4.63\times10^{-3}g_{11}B_1~{\rm Mpc}^{-1},\label{eq:dm}
\end{eqnarray}
where $m_a$ is the ALP mass, $\calE$ is the photon energy, $g$ is the
axion-photon interaction parameter. We adopt units such that
$c=\hbar=1$, and define dimensionless quantities
\begin{eqnarray}
m_1&=&m_a/(1\,{\rm neV}),\nonumber\\
\calE_1&=&\calE/(1\,{\rm TeV}),\nonumber\\
g_{11}&=&g/(10^{-11}{\rm GeV}^{-1}),\nonumber\\
B_1&=&B_{\rm tr}/(1\,{\rm nG}).
\end{eqnarray}
The plasma parameter $\Delta_{\rm pl}=-\omega_{\rm
  pl}^2/(2\omega)=-1.11\times 10^{-11}\calE_1^{-1} (n_e/10^{-7}{\rm
  cm}^{-3})~{\rm Mpc}^{-1}$ (where $\omega_{\rm pl}$ is the electron
plasma frequency and $n_e$ is the electron density) is unimportant for
the parameter regime considered in this paper and will be
neglected. Also, Eq.~(\ref{eq:eveq0}) does not include the QED effect,
which is negligible for typical nG intergalactic magnetic fields
~\cite{lai06}. All the numerical results presented in this paper are
based on Eq.~(\ref{eq:eveq0}).

\section{Analytical Results}

\begin{figure}[bt]
\includegraphics[angle=-90,width=1.\columnwidth]{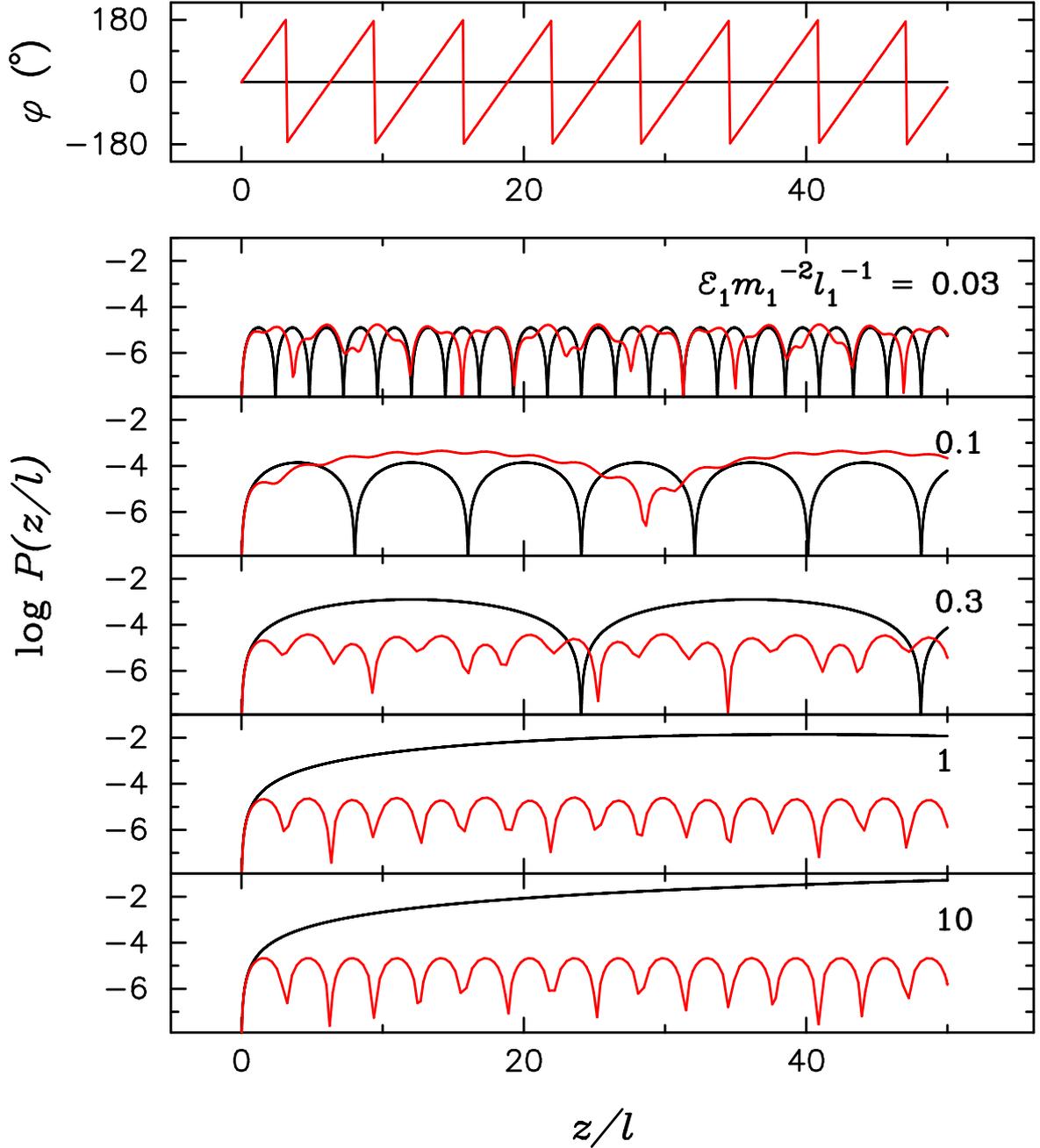}
\vspace{-0.5cm} \caption{Photon-to-ALP conversion probability in
  single magnetic domain models for various values of
  $\calE_1m_1^{-2}l_1^{-1}$, all with $g_{11}B_1l_1=1$. The black lines
  correspond to the model with $\varphi=0$ (constant field) in the
  whole domain along the ray, with $P(z)$ analytically given by
  Eq.~(\ref{eq:Pad}).  The red lines correspond to the model with
  $\varphi = l^{-1}z$, with $P(z)$ described by
  Eq.~(\ref{eq:P}).}
\label{fig:single-onedomain}
\end{figure}

For a given magnetic field structure, the ALP-photon evolution can
be obtained by integrating Eq.~(\ref{eq:eveq0}) along the ray.
Before studying complex random fields, we first consider two
simple ``single-domain'' cases: (i) $\varphi=0$ independent of
$z$; (ii) $\varphi$ increases linearly with $z$, with
$\varphi'=l^{-1}$. Typical intergalactic magnetic fields have a
coherence length of order $l\sim1\,$Mpc, we define
\begin{equation}
l_1=l/(1~{\rm Mpc}).
\end{equation}
Note that,  as a function of $z/l$, the photon-to-ALP conversion
probability $P$ depends only on the dimensionless quantities
$\Delta_al$ and $\Delta_M l$, and thus on
$\calE_1m_1^{-2}l_1^{-1}$ and $g_{11}B_1l_1$. Some numerical
results are plotted in Fig.~\ref{fig:single-onedomain}, showing
that the photon-to-ALP conversion probabilities are quite
different in the two cases.

To understand the difference analytically, we consider the evolution
of $E_\parallel$ and $E_\perp$, the components of $\vecE$ parallel and
perpendicular to $\vecB_{\rm tr}$, respectively.  Since
$E_\parallel=E_x\cos\varphi+E_y\sin\varphi$,
$E_\perp=-E_x\sin\varphi+E_y\cos\varphi$, Eq.~(\ref{eq:eveq0}) can be
rewritten as
\begin{equation}
i\lp\begin{array}{c} a'  \\ E_\parallel' \\ E_\perp'\\
\end{array}\rp = \lp \begin{array}{ccc}
\Delta_a  & \Delta_M & 0 \\
\Delta_M & 0 & i\varphi' \\
0 & -i\varphi'  & 0 \\ \end{array} \rp
\lp\begin{array}{c} a  \\ E_\parallel \\ E_\perp\\ \end{array}\rp,
\label{eq:eveq1}
\end{equation}
where we have dropped the non-essential term $\omega$ in the
diagonal elements.

If $|\varphi'|\ll |\Delta_M|$, Eq.~(\ref{eq:eveq1}) can be simplified
to the evolution equation of $a$ and $E_\parallel$:
\begin{equation}
i\lp\begin{array}{c} a'  \\ E_\parallel' \\
\end{array}\rp \simeq \lp \begin{array}{cc}
\Delta_a  & \Delta_M \\
\Delta_M & 0 \\ \end{array} \rp
\lp\begin{array}{c} a  \\ E_\parallel \\ \end{array}\rp.
\label{eq:eveq-ad}
\end{equation}
This equation has been widely discussed in previous works.  If the
magnetic field strength varies slowly (we assume $B_{\rm tr}$ is
constant along the ray for simplicity), the mode evolution is said to
be ``adiabatic'', and the photon-to-ALP conversion probability is
given by the well-known formula~\cite{kuo89}:
\begin{equation}
P_{\rm ad}=\frac{\Delta_M^2}{(\Delta k/2)^2} \sin^2(\Delta
kz/2),\label{eq:Pad}
\end{equation}
with $\Delta k=\sqrt{\Delta_a^2+4\Delta_M^2}$. In the limit of $\Delta
kz/2 \ll 1$, Eq.~(\ref{eq:Pad}) simplifies to $P_{\rm ad}\simeq
\Delta_M^2z^2$.

Intergalactic magnetic fields can often have $|\varphi'|\sim 1\,{\rm
  Mpc}^{-1}$, much larger than $|\Delta_a|$ and $\Delta_M$ (see
Eqs.~2-3). If $|a|\ll|E|$, the electric field can be solved as
$E_\parallel\simeq\cos\varphi$, $E_\perp\simeq\sin\varphi$ assuming
$E_x=1$, $E_y=0$ at $z=0$, i.e., $\vecE(z)\simeq
\vecE(z=0)$. Substitute this electric field into Eq.~(\ref{eq:eveq1}),
we find the evolution equation for the ALP field, $ia'\simeq\Delta_a
a+\Delta_M\cos\varphi$, with the solution
\begin{equation}
a(z)\simeq-e^{-i\Delta_az}i\int_0^z\intd z \Delta_M\cos\varphi(z) e^{i\Delta_a z}. \label{eq:a}
\end{equation}
For $\varphi(z)=\varphi'z$ with constant $\varphi'$,
we obtain the photon-to-ALP conversion probability
\begin{gather}
P=|a(z)|^2\simeq\frac{\Delta_M^2}{\Delta_a^2(1-\varphi'^2/\Delta_a^2)^2}
\bigl[(\cos\varphi-\cos\Delta_az)^2
\nonumber\\
+(\varphi' \sin\varphi/\Delta_a-\sin\Delta_az)^2\bigr].
\label{eq:P}
\end{gather}
This equation accurately describes the numerical result of
Fig.~\ref{fig:single-onedomain} for various values of $\calE
m_a^{-2}$. For example, in the limit of $|\varphi'|\gg|\Delta_a|$,
Eq.~(\ref{eq:P}) simplifies to $P\simeq
(\Delta_M/\varphi')^2\sin^2\varphi$, which has an oscillation length
$\pi/\varphi'$ and is independent of the ALP mass and energy.

\section{Results for Random Magnetic Fields}

\begin{figure}[bt]
\includegraphics[angle=-90,width=1.\columnwidth]{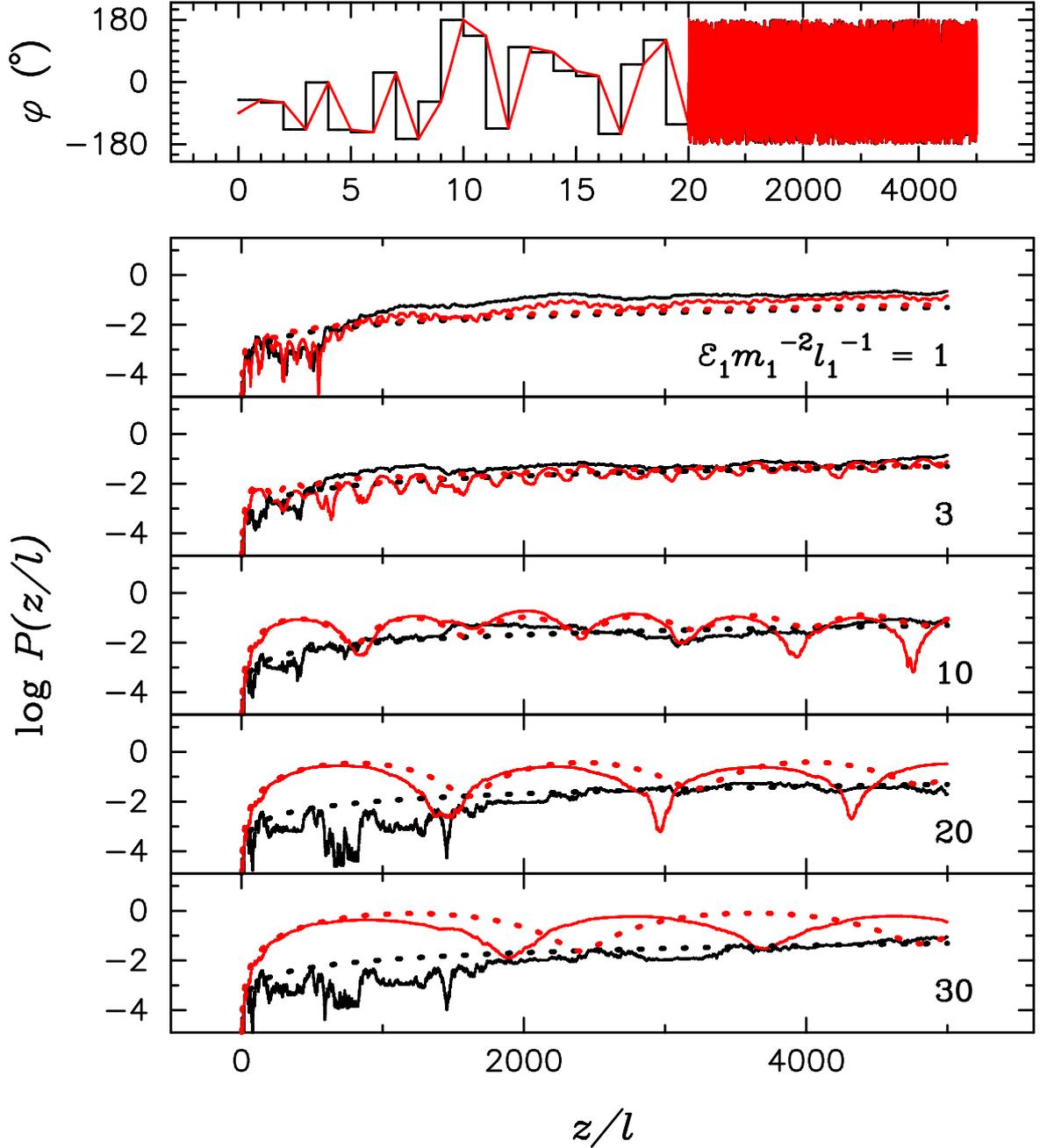}
\vspace{-0.5cm} \caption{Photon-to-ALP conversion probability across
  multiple domains of the intergalactic medium with random magnetic
  fields.  Each domain has the same size $l$ and magnetic field
  strength.  The top panel depicts an example of the magnetic
  orientation angle in two different models: The black line for the
  discrete-$\varphi$ model and the red line for the
  linearly-continuous-$\varphi$ model.  The lower panels show the
  numerical results for the conversion probability for various values
  of $\calE m_a^{-2}l^{-1}$ (all with $B_1g_{11}l_1=1$), for the
  discrete-$\varphi$ model (black lines) and the
  linearly-continuous-$\varphi$ model (red lines).  The black-dotted
  and red-dotted lines correspond to the analytical expressions
  (\ref{eq:PadN}) and (\ref{eq:PN}),
  respectively. } \label{fig:single}
\end{figure}

The magnetic field in the intergalactic medium is randomly
distributed, with the expected coherent length of order 0.1-1~Mpc
(about the size of galaxy clusters). In general, numerical
integrations are necessary to obtain the photon-to-axion
conversion probability for a given realization of the random
magnetic field distributions in addition to the relevant ALP
parameters. A ``discrete-$\varphi$'' model has been widely used in
previous studies: The path of propagation is divided into many
domains, each has the same size $l$ and a uniform magnetic field,
with the magnetic orientation angle $\varphi$ changing randomly
but discretely from one domain to the next. Based on this model,
Ref.~\cite{gro02} derived an analytic expression for the mean
value of the photo-to-ALP conversion probability after propagating
through $N$ domains (over distance $z= Nl$):
\begin{equation}
P_{{\rm ad},N}=\frac{1}{3}\left(1-e^{-3NP_{\rm ad}/2}\right), \label{eq:PadN}
\end{equation}
where on the right-hand side, $P_{\rm ad}$ is given by
Eq.~(\ref{eq:Pad}) evaluated at $z=l$. Note that for $NP_{\rm
  ad}\gg1$, we have $P_{{\rm ad},N}=1/3$, an upper limit for the
conversion probability. In Fig.~\ref{fig:single} we depict an
example of the discrete-$\varphi$ model and the numerical results
for the conversion probabilities at different values of $\calE
m_a^{-2}$ (black lines). These numerical results are in agreement
with Eq.~(\ref{eq:PadN}) in the statistical sense.

As discussed above, we expect that the discrete-$\varphi$ model
may be problematic since in most regions of the intergalactic
medium $|\varphi'|$ can be much larger than $\Delta_M$. In
Fig.~\ref{fig:single} (see the red lines) we consider a
``linearly-continuous-$\varphi$'' model: The path of propagation
is again divided into many equal-sized domains; in each domain,
$\varphi$ varies linearly from one random value to another (thus,
$\varphi$ is always continuous, $\varphi'$ is constant inside each
domain but changes across the domain boundary).  Our numerical
results show that this continuous-$\varphi$ model can yield
completely different conversion probabilities compared to the
discrete-$\varphi$ model. In particular, $P(z/l)$ exhibits
quasi-periodicity along the ray (with the period dependent on
$\calE m_a^{-2}l^{-1}$) and can be close to unity for large values
of $\calE_1 m_1^{-2}l_1^{-1}$.

To understand these numerical results, we apply Eq.~(\ref{eq:a}) to
the linearly-continuous-$\varphi$ model. The ALP amplitude after
traversing $N$ domains is given by
\begin{equation}
a_N\simeq
-e^{-iN\Delta_al}i\sum_{j=1}^{N}\Delta_M\int_{(j-1)l}^{jl}
\cos\varphi(z) e^{i\Delta_a z}\intd z. \label{eq:aN}
\end{equation}
In the $j$-th domain, $\varphi(z)=\varphi_{j-1}+\varphi'[z-(j-1)l]$,
with $\varphi'=(\varphi_j-\varphi_{j-1})/l$.  For
$|\varphi'|\gg|\Delta_a|$ and $|\Delta_a| l\ll1$ (these two conditions
are similar since $|\varphi'|\sim l^{-1}$), Eq.~(\ref{eq:aN}) can be
simplified, giving
\begin{equation}
|a_N|^2\simeq \Delta_M^2l^2\biggl|\sum_{j=1}^{N}
A_j(\varphi_j,\varphi_{j-1})e^{ij\Delta_al}\biggr|^2,
\end{equation}
where
\begin{gather}
A_j\simeq\frac{\Delta\varphi(\sin\varphi_j-\sin\varphi_{j-1})}{\Delta\varphi^2-\Delta_a^2l^2}
+\frac{i\Delta_al(\cos\varphi_j-\cos\varphi_{j-1})}{\Delta\varphi^2-\Delta_a^2l^2},
\label{eq:A}
\end{gather}
with $\Delta\varphi=\varphi'l=\varphi_j-\varphi_{j-1}$. For random
$\varphi_j$ (varying between $-\pi$ and $\pi$), $A_j$ can be
characterized by the mean $\langle A\rangle$ and variance
$\sigma_A^2=\langle |A_j-\langle A\rangle|^2\rangle$.  The mean
photo-to-ALP conversion probability $P_N=\langle |a_N|^2\rangle$ is
then
\begin{equation}
P_N\simeq  0.123\frac{\Delta_M^2}{\Delta_a^2}(1-\cos
N\Delta_al)+\sigma_A^2N\Delta_M^2l^2, \label{eq:PN}
\end{equation}
where we have used $|\langle A\rangle|\simeq
\langle\frac{\sin\varphi_j-\sin\varphi_{j-1}}
{\varphi_i-\varphi_{j-1}}\rangle\simeq 0.248$. The variance
$\sigma_A^2$ can be calculated using Monte-Carlo method, and we
find $\sigma_A^2\simeq 0.44,\,0.30$ and $0.23$ for
$\calE_1m_1^{-2}l_1^{-1}=0.3,\,1$ and $\go 10$ (corresponding to
$|\Delta_a| l=0.26,\,0.078$ and $\lo 0.0078$). Note that the
validity of Eq.~(\ref{eq:PN}) requires $|\Delta_a| l\ll 1$,
$\Delta_M l\ll 1$ and $|a|\ll |\vecE|$ (or $P_N\ll 1$). Under the
same condition, Eq.~(\ref{eq:PadN}) reduces to $P_{\rm ad,N}\simeq
0.5 N\Delta_M^2 l^2$, similar to the second term in
Eq.~(\ref{eq:PN}).

Equation~(\ref{eq:PN}) indicates that the photon-to-ALP conversion
probability has a cosine function dependence, with the oscillation
length (in units of $l$) $2\pi/|\Delta_al| \simeq
80\,\calE_1m_1^{-2}l_1^{-1}$. This is in agreement with numerical
results presented in Fig.~\ref{fig:single}, especially for $1\lo
\calE_1m_1^{-2}l_1^{-1}\lo 30$.  For $\calE_1m_1^{-2}l_1^{-1}\lo
1$, the inequality $|\Delta_a|\ll|\varphi'|$ is not well
satisfied; for $\calE_1m_1^{-2}l_1^{-1}\go 30$, the ALP amplitude
can be comparable to $|\vecE|$, making Eq.~(\ref{eq:PN})
inaccurate.

\section{Distribution of Conversion Probability}

\begin{figure}[bt]
\includegraphics[angle=-90,width=1.\columnwidth]{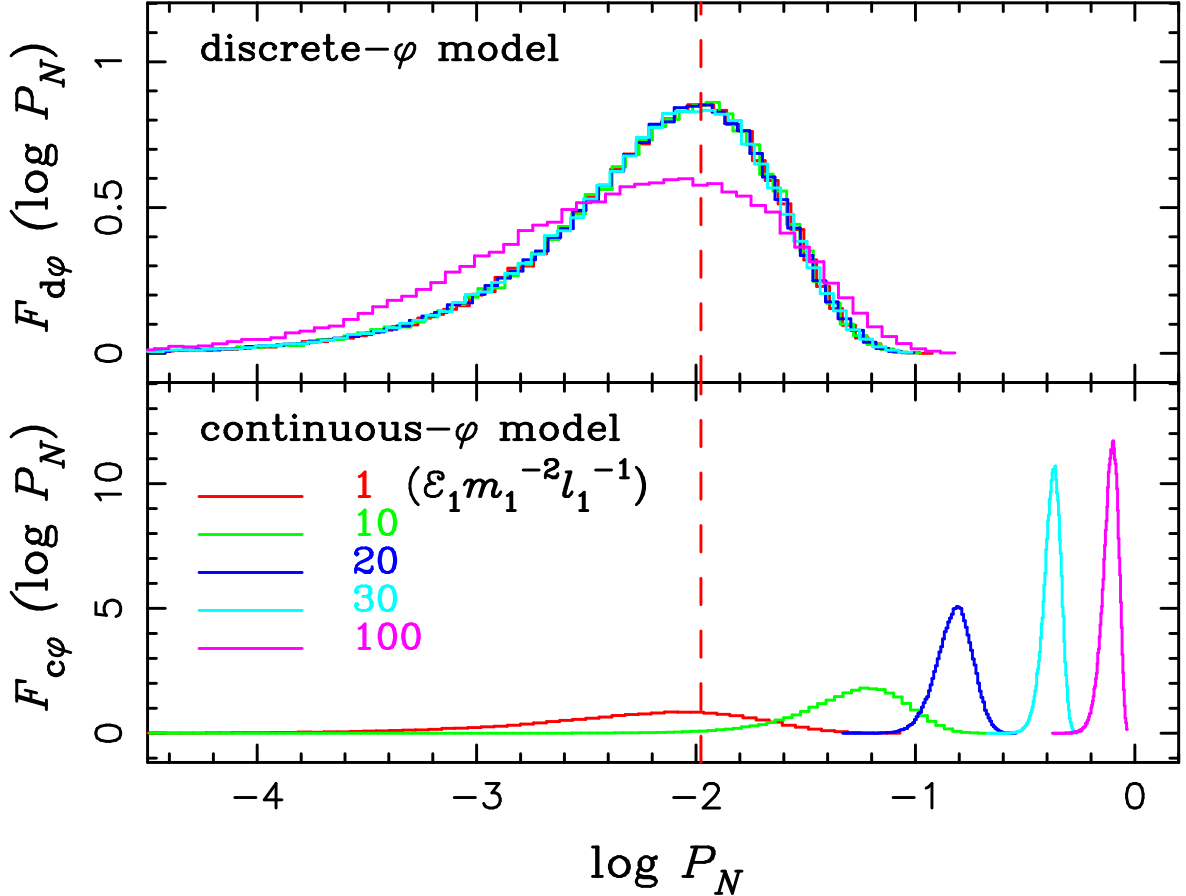}
\vspace{-0.5cm} \caption{The distribution function of the
  photon-to-ALP conversion probability after a distance of 1~Gpc.  The
  upper and lower panels correspond to the discrete-$\varphi$ model
  and the linearly-continuous-$\varphi$ model, respectively (see
  Fig.~\ref{fig:single}).  The different curves are for different
  values of $\calE m_a^{-2}l^{-1}$, all with $g_{11}B_1l_1^{-1}=1$, the domain size
  $l=1$~Mpc and the domain number $N=1000$.  The vertical red dashed line
  represents the theoretical conversion probability of the
  discrete-$\varphi$ model, given by Eq.~(\ref{eq:PadN}).
} \label{fig:fPN}
\end{figure}

Because the intergalactic magnetic field has random orientations
along the propagation path, the conversion probability $P_N$ has a
distribution with finite spread.  To obtain the $P_N$
distribution, we carry out Monte-Carlo calculations of the
photon-ALP propagations for $10^5$ times, each time with the same
set of ALP and magnetic field parameters, but for different random
values of $\varphi$ in each domain.  We consider both the discrete
$\varphi$ model and the linearly-continuous $\varphi$ model as
discussed above. The results are shown in Fig.~\ref{fig:fPN}.

For the discrete-$\varphi$ model, the $P_N$-distribution function,
$F_{d\varphi}(\log P_N)$, is a skewed Gaussian (see the upper
panel of Fig.~\ref{fig:fPN}). The peak of the distribution is
accurately predicted by Eq.~(\ref{eq:PadN}).  We find that
$F_{d\varphi}(\log P_N)$ is almost the same for different values
of $\calE m_a^{-2}l^{-1}$, except that for
$\calE_1m_1^{-2}l_1^{-1}\go 40$ the distribution becomes a
broader.

The lower panel of Fig.~\ref{fig:fPN} shows the $P_N$-distribution
function $F_{c\varphi}(\log P_N)$ for the linearly-continuous
$\varphi$ model.  For $\calE_1m_1^{-2}l_1^{-1}\lo 1$ the
distribution function is similar to that of the discrete-$\phi$
model. However, as $\calE_1m_1^{-2}l_1^{-1}$ increases, the peak
of the distribution shifts to larger values and the width becomes
narrower -- these features are in marked contrast to the
discrete-$\varphi$ model.

\begin{figure}[bt]
\includegraphics[angle=-90,width=1.\columnwidth]{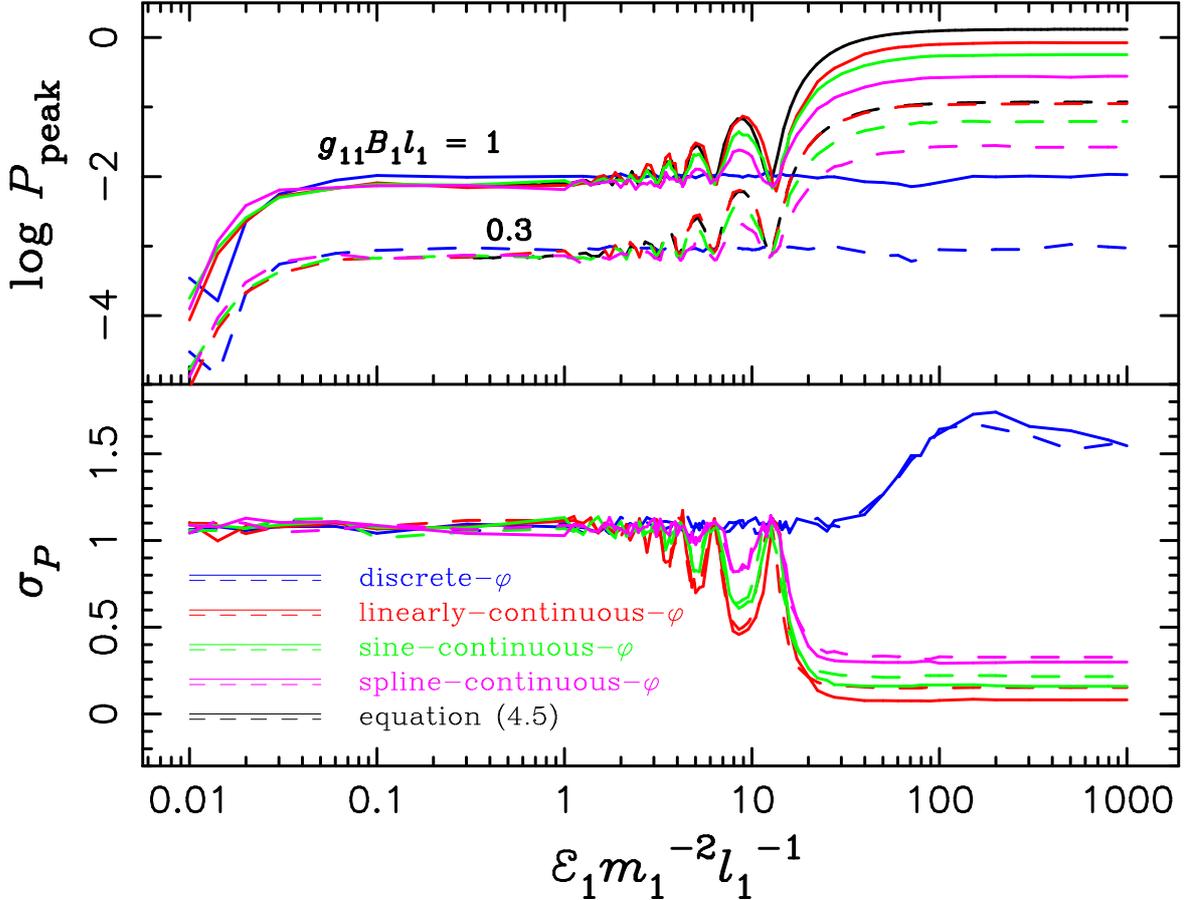}
\vspace{-0.5cm} \caption{Peak position $P_{\rm peak}$ and half-peak
  width $\sigma_P$ of the $P_N$-distribution function across $\calE
  m_a^{-2}l^{-1}$. Two different values of $g_{11}B_1l_1$ ($=0.3,\,1$)
  are presented as dashed and solid lines. Lines with different colors
  correspond to different $\varphi$ models (as indicated
  in the lower panel). The black lines in the upper panel show the
  analytical equation~(\ref{eq:PN}).  The other parameters are the
  same as in Fig.~\ref{fig:fPN}.}
\label{fig:Ppeak}
\end{figure}

\begin{figure}[bt]
\begin{tabular}{cc}
\includegraphics[angle=-90,width=0.45\columnwidth]{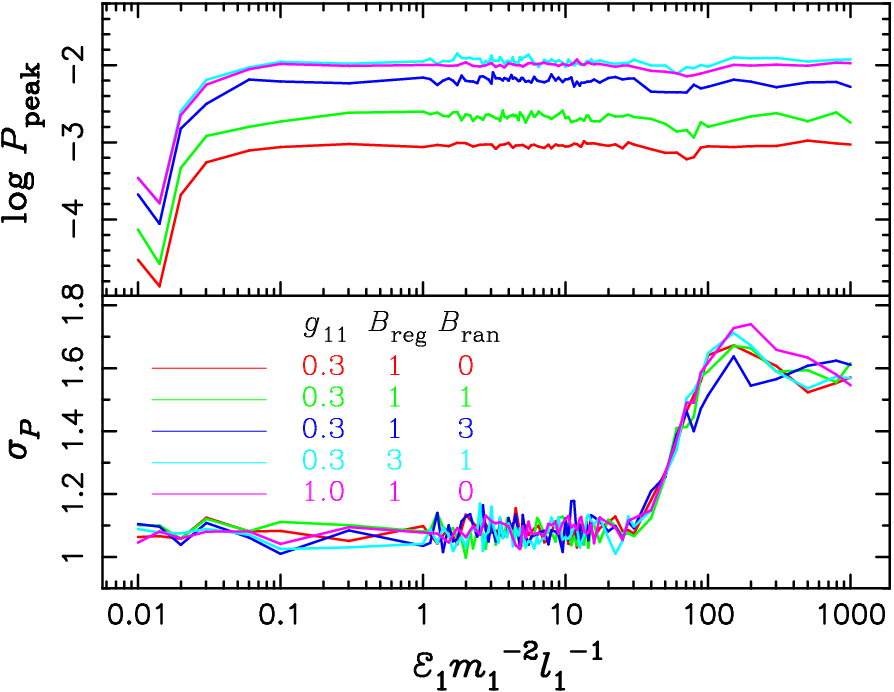} &
\includegraphics[angle=-90,width=0.45\columnwidth]{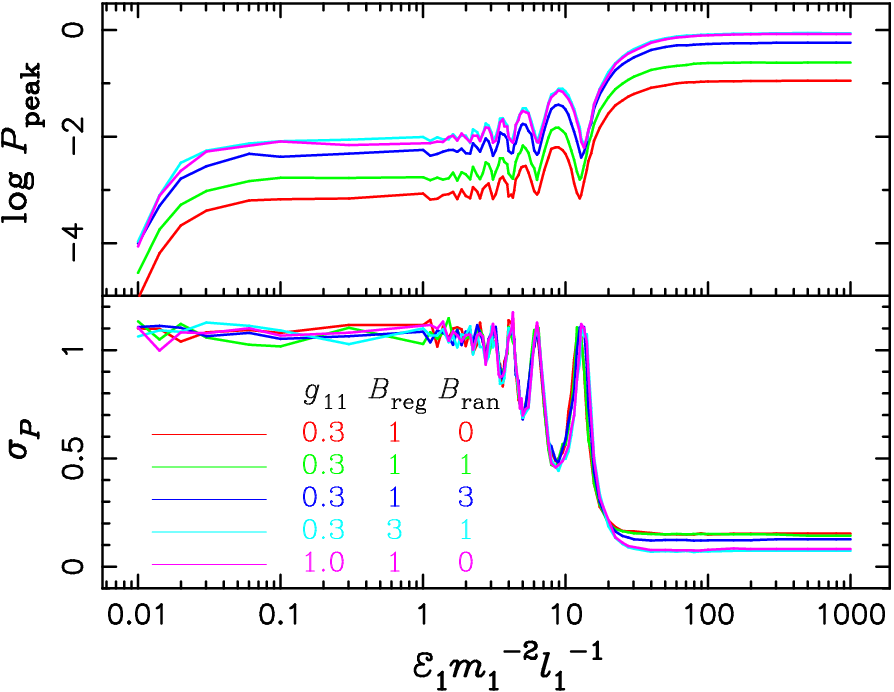}\\
a) random $B_{\rm tr}$ with discrete-$\varphi$ model&
b) random $B_{\rm tr}$ with linearly-continuous-$\varphi$ model \\
\end{tabular}
\vspace{-0.5cm} \caption{Similar to Fig.~\ref{fig:Ppeak}, except that
  the magnetic field magnitudes $B_{\rm tr}$ in different domains are
  different and are randomly distributed in the range of $B_{\rm reg}$
  and $B_{\rm reg}+B_{\rm ran}$.  The Colored lines correspond to
  different values of $g_{11}$, $B_{\rm reg}$ and $B_{\rm ran}$ as
  indicated (the units of $B$ is nG). Panel a) and b) represent the
  discrete-$\varphi$ and linearly-continuous-$\varphi$ models,
  respectively. The other parameters are the same as in
  Fig.~\ref{fig:fPN}.}
\label{fig:Ppeak-ranB}
\end{figure}

To characterize how the $P_N$-distribution function varies for
different parameters, we show in Fig.~\ref{fig:Ppeak} $P_{\rm peak}$
and $\sigma_P$, the peak and half-peak width of the distribution for
two different values of $g_{11}B_1l_1$, as a function of
$\calE_1m_1^{-2}l_1^{-1}$ (the other parameters are the same as in
Fig.~\ref{fig:fPN}).  For the discrete-$\varphi$ model (blue lines in
Fig.~\ref{fig:Ppeak}), $P_{\rm peak}$ is almost independent of
$\calE_1m_1^{-2}l_1^{-1}$, except when the oscillation length is
smaller than domain size, i.e., when $\Delta kl/2\go 1$ or
$\calE_1m_1^{-2}l_1^{-1}\lo 0.04$. The value of $P_{\rm peak}$ can be
accurately predicted by Eq.~(\ref{eq:PadN}). The width of the
distribution is almost constant except for
$\calE_1m_1^{-2}l_1^{-1}\go40$. Note that for $g_{11}B_1l_1\gtrsim
10$, the conversion probability is close to the upper limit 1/3, and
the $P_N$ distribution is not a Gaussian.

For the linearly-continuous $\varphi$ model (red lines in
Fig.~\ref{fig:Ppeak}), larger $\calE_1m_1^{-2}l_1^{-1}$ generally
leads to larger $P_{\rm peak}$ and smaller $\sigma_P$. Interestingly,
both $P_{\rm peak}$ and $\sigma_P$ are not a monotonous function of
$\calE_1m_1^{-2}l_1^{-1}$, but have oscillations. This oscillation can
be described by Eq.~(\ref{eq:PN}), as shown by the black lines in
Fig.~\ref{fig:Ppeak} [Note that Eq.~(\ref{eq:PN}) is valid only for
  $|\Delta_a|l\ll 1$, so we choose the dotted lines start from
  $\calE_1m_1^{-2}l_1^{-1}\simeq0.3$].  For
$\calE_1m_1^{-2}l_1^{-1}\lo1$, both $P_{\rm peak}$ and $\sigma_P$ are
almost the same as in the discrete-$\varphi$ model. In the case of
$g_{11}B_1l_1=0.3$ (blue lines), Eq.~(\ref{eq:PN}) agrees very well
with the numerical result, since the assumption $|a|\ll |\vecE|$
always tenable.  For $|N\Delta_al | \ll 1$ [or
  $\calE_1m_1^{-2}l_1^{-1}\gg 78N/10^3$], the conversion probability
reaches its maximum $P_{\rm max}\simeq 0.0615N^2\Delta_M^2l^2 = 1.3
(g_{11}B_1l_1)^2(N/10^3)^2$. In the case of $g_{11}B_1l_1=1$, the peak
conversion probability $P_{\rm peak}$ approaches unity for
$\calE_1m_1^{-2}l_1^{-1}\go 40$, implying a nearly $100\%$
photon-to-ALP conversion [Of course, the analytical expression
  (\ref{eq:PN}) is less accurate when $P_{\rm peak}\sim 1$ since
  $|a|\ll |\vecE|$ is invalid].

In the above, we have focused on the linearly-continuous-$\varphi$
model, since in this case we can derive analytical equations [see
  Eqs.~(\ref{eq:P}) and (\ref{eq:PN})] to help understand our
numerical results.  We have performed calculations for other
continuous-$\varphi$ models, e.g., using the spline function or sine
function to link the random $\varphi$ values in multiple domains (see
Fig.~\ref{fig:Ppeak}).  We find that the results for the $P_N$
distribution are similar to the linearly-continuous-$\varphi$ model,
although the conversion probabilities are slightly lower because in
the spline and sine $\varphi$ models there always exist some regions
with $\varphi'\sim 0$.

So far in this paper we have assumed that the magnetic field has the
same strength $B_{\rm tr}$ in different domains but with varying
orientations.  What happens when $B_{\rm tr}$ also varies?  For
concreteness, we consider a simple model where the values of $B_{\rm
  tr}$ in different domains are randomly distributed in the range
between $B_{\rm reg}$ and $B_{\rm reg}+B_{\rm ran}$. Our numerical
results for the final $P_N$ distributions (for both discrete-$\varphi$
and continuous-$\varphi$ models) for various values of $g$, $B_{\rm
  reg}$ and $B_{\rm ran}$ are shown in
Figure~\ref{fig:Ppeak-ranB}. The results are very similar to the
constant $B_{\rm tr}$ case shown in Fig.~\ref{fig:Ppeak}. For the
discrete-$\varphi$ model (Fig.~\ref{fig:Ppeak-ranB}a), we can derive
an analytical expression of the final conversion probability using
Eq.~(\ref{eq:aN}) (with $\varphi(z)=$constant in each domain, but
$\Delta_M$ and $\varphi$ have different values in different domains).
For $|\Delta _a| l\ll 1$, we find
\begin{equation}
P_{{\rm ranB},N}=N\left<\Delta_M^2l^2\cos^2\varphi_j\right>
  = \frac{1}{8}Ng^2l^2\left<B_{\rm tr}^2\right>, \label{eq:PranB}
\end{equation}
where $\left<B_{\rm tr}^2\right>=B_{\rm reg}^2+B_{\rm reg}B_{\rm
  ran}+B_{\rm ran}^2/3$.  The above equation is the same as the case
with constant $B_{\rm tr}$ and discrete $\varphi$ ($P_{{\rm
    ad},N}=0.5N\Delta_M^2l^2$), if we replace $B_{\rm tr}$ by
$\sqrt{\left<B_{\rm tr}^2\right>}=\sqrt{B_{\rm reg}^2+B_{\rm
    reg}B_{\rm ran}+B_{\rm ran}^2/3}$. Equation~(\ref{eq:PranB})
agrees well with the numerical results shown in
Fig.~\ref{fig:Ppeak-ranB}a.  The same applies for the
linearly-continuous-$\varphi$ model: the $P_{\rm peak}$ curves shown
in Fig.~\ref{fig:Ppeak-ranB}b are almost the same as the constant
$B_{\rm tr}$ case if we replace $B_{\rm tr}$ by $\sqrt{\left<B_{\rm
    tr}^2\right>}$.

\section{Discussion}

We have shown that a proper treatment of the inhomogeneity of
intergalactic magnetic fields can lead to very different
photon-to-ALP conversion probabilities compared to the
``discrete-$\varphi$'' model widely used in previous studies. The
difference is particularly striking when
$\calE_1m_1^{-2}l_1^{-1}\go 4 \sqrt{N/10^3}$ [the first term of
Eq.~(\ref{eq:PN}) larger than the second term; here $l$ is the
coherence length of the magnetic field domain and $N=z/l$ the
domain numbers across distance $z$]. In the discrete-$\varphi$
model, the conversion probability is determined by $\Delta_Ml
\propto gBl$ and almost does not depend on $\Delta_al$ or $\calE
m_a^{-2}l^{-1}$ (assuming $|\Delta_a| l\ll 1$ and $\Delta_M l\ll
1$), and never exceeds $1/3$ [see Eq.~(\ref{eq:PadN})]. By
contrast, in the continuous-$\varphi$ model, the photon-to-axion
conversion probability has a distinct dependence on $\calE_1
m_1^{-2}l_1^{-1}$; it becomes significant when $|\Delta_a|\lo
\Delta_M$ [see Eqs.~(2)-(3)] and can be as large as 100\% (see
Figs.~\ref{fig:fPN}-\ref{fig:Ppeak}).  Our analytic expression
(\ref{eq:PN}) (valid for $|\Delta_a| l\ll 1$, $\Delta_M l\ll 1$
and $P_N\lo 1$) approximately captures these features.

Note that although we have considered TeV photon-ALP propagation in
intergalactic magnetic fields (with $B_{\rm tr}\sim 1\,{\rm nG}$,
$l\sim 1\,{\rm Mpc}$), our results can be easily re-scaled to
different situations. For example, Ref.~\cite{hor12} explored the
hardening of the TeV photon spectrum of Blazars due to the $\gamma\to
a\to\gamma$ conversions in the magnetic fields of galaxy clusters and
Milky Way. In galaxy clusters, $B_{\rm tr}\sim1\,{\rm \mu G}$,
$l\sim10\,{\rm kpc}$, we have
\begin{eqnarray}
\calE_1m_1^{-2}l_1^{-1}&=&\frac{\calE}{1\,{\rm
TeV}}\left(\frac{m_a}{10^{-8}\,{\rm
eV}}\right)^{-2}\left(\frac{l}{10\,{\rm kpc}}\right)^{-1},\nonumber\\
g_{11}B_1l_1&=&\frac{g}{10^{-12}\,{\rm GeV}}\frac{B_{\rm tr}}{1\,{\rm \mu
G}}\frac{l}{10\,{\rm kpc}}.
\end{eqnarray}
Thus with $m_a\sim10^{-8}\,$eV, $g\sim10^{-12}\,$GeV (see
\cite{hor12}), the conversion of TeV photon-to-ALP in the galaxy
clusters can be significantly affected by our results. ALPs can
convert back to be TeV photons in the magnetic field of the Milky
Way. Similar ALP-to-photon conversion in M31 and the Milky Way are
used to explain the recent observations of the 3.55~keV photon
line~\cite{hig14,jae14,cic14,aba14,con14}. Our results can be easily
adapted to the typical galactic magnetic field ($B_{\rm tr}\sim
10\,{\rm \mu G}$, $l\sim 100\,$pc--$1\,$kpc) with different $\calE$,
$g$ and $m_a$.

In summary, many previous works use the discrete-$\varphi$ model
and Eq.~(\ref{eq:PadN}) to estimate the photon-to-ALP conversion
probabilities. In light of the significant difference between the
discrete and continuous-$\varphi$ models of magnetic fields, a
re-evaluation of the previous results is warranted.

\section*{Acknowledgements}
This work has been supported in part by the National Natural
Science Foundation of China (11273029), and by NSF grant
AST-1211061 and NASA grant NNX14AG94G.

\vspace{-0.5cm}
\bibliographystyle{plain}

\bibliography{ms}

\begin{thebibliography}{10}

\bibitem{pq77}
R.~D. Peccei and Helen~R. Quinn.
\newblock Cp conservation in the presence of pseudoparticles.
\newblock {\em Physical Review Letters}, 38:1440--1443, 1977.

\bibitem{jae10}
Joerg Jaeckel and Andreas Ringwald.
\newblock The low-energy frontier of particle physics.
\newblock {\em Annual Review of Nuclear and Particle Science}, 60:405--437,
  2010.

\bibitem{mir08}
Alessandro Mirizzi, Georg~G. Raffelt1, Pasquale~D. Serpico, Georg Raffelt, and
  Berta Beltrán.
\newblock Photon-axion conversion in intergalactic magnetic fields and
  cosmological consequences, n/a 1, 2008 2008.

\bibitem{csa02}
Csaba Csáki, Nemanja Kaloper, and John Terning.
\newblock Dimming supernovae without cosmic acceleration.
\newblock {\em Physical Review Letters}, 88:161302, 2002.

\bibitem{mir05}
Alessandro Mirizzi, Georg~G. Raffelt, and Pasquale~D. Serpico.
\newblock Photon-axion conversion as a mechanism for supernova dimming: Limits
  from cmb spectral distortion.
\newblock {\em Physical Review D}, 72:23501, 2005.

\bibitem{ost05}
Linda Östman and Edvard Mörtsell.
\newblock Limiting the dimming of distant type ia supernovae.
\newblock {\em Journal of Cosmology and Astro-Particle Physics}, 02:005, 2005.

\bibitem{dia14}
A.~G. Dias, A.~C.~B. Machado, C.~C. Nishi, A.~Ringwald, and P.~Vaudrevange.
\newblock The quest for an intermediate-scale accidental axion and further
  alps.
\newblock {\em Journal of High Energy Physics}, 06, 2014.

\bibitem{csa14}
Csaba Csaki, Nemanja Kaloper, and John Terning.
\newblock Planck data and ultralight axions, May 1, 2014 2014.
\newblock 15 pages, 4 figures.

\bibitem{aha06}
F.~Aharonian, A.~G. Akhperjanian, A.~R. Bazer-Bachi, M.~Beilicke, W.~Benbow,
  D.~Berge, K.~Bernlöhr, C.~Boisson, O.~Bolz, V.~Borrel, I.~Braun,
  F.~Breitling, A.~M. Brown, P.~M. Chadwick, L.-M. Chounet, R.~Cornils,
  L.~Costamante, B.~Degrange, H.~J. Dickinson, A.~Djannati-Ataï, L.~O'c.
  Drury, G.~Dubus, D.~Emmanoulopoulos, P.~Espigat, F.~Feinstein, G.~Fontaine,
  Y.~Fuchs, S.~Funk, Y.~A. Gallant, B.~Giebels, S.~Gillessen, J.~F.
  Glicenstein, P.~Goret, C.~Hadjichristidis, D.~Hauser, M.~Hauser,
  G.~Heinzelmann, G.~Henri, G.~Hermann, J.~A. Hinton, W.~Hofmann, M.~Holleran,
  D.~Horns, A.~Jacholkowska, O.~C. de~Jager, B.~Khélifi, S.~Klages, Nu. Komin,
  A.~Konopelko, I.~J. Latham, R.~Le~Gallou, A.~Lemière, M.~Lemoine-Goumard,
  N.~Leroy, T.~Lohse, J.~M. Martin, O.~Martineau-Huynh, A.~Marcowith,
  C.~Masterson, T.~J.~L. McComb, M.~de~Naurois, S.~J. Nolan, A.~Noutsos, K.~J.
  Orford, J.~L. Osborne, M.~Ouchrif, M.~Panter, G.~Pelletier, S.~Pita,
  G.~Pühlhofer, M.~Punch, B.~C. Raubenheimer, M.~Raue, J.~Raux, S.~M. Rayner,
  A.~Reimer, O.~Reimer, J.~Ripken, L.~Rob, L.~Rolland, G.~Rowell, V.~Sahakian,
  L.~Saugé, S.~Schlenker, R.~Schlickeiser, C.~Schuster, U.~Schwanke,
  M.~Siewert, H.~Sol, D.~Spangler, R.~Steenkamp, C.~Stegmann, J.-P. Tavernet,
  R.~Terrier, C.~G. Théoret, M.~Tluczykont, C.~van Eldik, G.~Vasileiadis,
  C.~Venter, P.~Vincent, et~al.
\newblock A low level of extragalactic background light as revealed by γ-rays
  from blazars.
\newblock {\em Nature}, 440:1018--1021, 2006.

\bibitem{maz07}
D.~Mazin and M.~Raue.
\newblock New limits on the density of the extragalactic background light in
  the optical to the far infrared from the spectra of all known tev blazars.
\newblock {\em Astronomy and Astrophysics}, 471:439--452, 2007.

\bibitem{alb08}
MAGIC Collaboration, J.~Albert, E.~Aliu, H.~Anderhub, L.~A. Antonelli,
  P.~Antoranz, M.~Backes, C.~Baixeras, J.~A. Barrio, H.~Bartko, D.~Bastieri,
  J.~K. Becker, W.~Bednarek, K.~Berger, E.~Bernardini, C.~Bigongiari,
  A.~Biland, R.~K. Bock, G.~Bonnoli, P.~Bordas, V.~Bosch-Ramon, T.~Bretz,
  I.~Britvitch, M.~Camara, E.~Carmona, A.~Chilingarian, S.~Commichau, J.~L.
  Contreras, J.~Cortina, M.~T. Costado, S.~Covino, V.~Curtef, F.~Dazzi,
  A.~De~Angelis, E.~De Cea~del Pozo, R.~de~los Reyes, B.~De~Lotto, M.~De~Maria,
  F.~De~Sabata, C.~Delgado Mendez, A.~Dominguez, D.~Dorner, M.~Doro,
  M.~Errando, M.~Fagiolini, D.~Ferenc, E.~Fernández, R.~Firpo, M.~V. Fonseca,
  L.~Font, N.~Galante, R.~J. García~López, M.~Garczarczyk, M.~Gaug,
  F.~Goebel, M.~Hayashida, A.~Herrero, D.~Höhne, J.~Hose, C.~C. Hsu, S.~Huber,
  T.~Jogler, T.~M. Kneiske, D.~Kranich, A.~La~Barbera, A.~Laille, E.~Leonardo,
  E.~Lindfors, S.~Lombardi, F.~Longo, M.~López, E.~Lorenz, P.~Majumdar,
  G.~Maneva, N.~Mankuzhiyil, K.~Mannheim, L.~Maraschi, M.~Mariotti,
  M.~Martínez, D.~Mazin, M.~Meucci, M.~Meyer, J.~M. Miranda, R.~Mirzoyan,
  S.~Mizobuchi, M.~Moles, A.~Moralejo, D.~Nieto, K.~Nilsson, J.~Ninkovic,
  N.~Otte, I.~Oya, M.~Panniello, R.~Paoletti, J.~M. Paredes, M.~Pasanen,
  D.~Pascoli, F.~Pauss, R.~G. Pegna, M.~A. Perez-Torres, et~al.
\newblock Very-high-energy gamma rays from a distant quasar: How transparent is
  the universe?
\newblock {\em Science}, 320:1752--, 2008.

\bibitem{ack12}
M.~Ackermann, M.~Ajello, A.~Allafort, P.~Schady, L.~Baldini, J.~Ballet,
  G.~Barbiellini, D.~Bastieri, R.~Bellazzini, R.~D. Blandford, E.~D. Bloom,
  A.~W. Borgland, E.~Bottacini, A.~Bouvier, J.~Bregeon, M.~Brigida, P.~Bruel,
  R.~Buehler, S.~Buson, G.~A. Caliandro, R.~A. Cameron, P.~A. Caraveo,
  E.~Cavazzuti, C.~Cecchi, E.~Charles, R.~C.~G. Chaves, A.~Chekhtman, C.~C.
  Cheung, J.~Chiang, G.~Chiaro, S.~Ciprini, R.~Claus, J.~Cohen-Tanugi,
  J.~Conrad, S.~Cutini, F.~D'Ammando, F.~de~Palma, C.~D. Dermer, S.~W. Digel,
  E.~do~Couto~e Silva, A.~Domínguez, P.~S. Drell, A.~Drlica-Wagner,
  C.~Favuzzi, S.~J. Fegan, W.~B. Focke, A.~Franckowiak, Y.~Fukazawa, S.~Funk,
  P.~Fusco, F.~Gargano, D.~Gasparrini, N.~Gehrels, S.~Germani, N.~Giglietto,
  F.~Giordano, M.~Giroletti, T.~Glanzman, G.~Godfrey, I.~A. Grenier, J.~E.
  Grove, S.~Guiriec, M.~Gustafsson, D.~Hadasch, M.~Hayashida, E.~Hays, M.~S.
  Jackson, T.~Jogler, J.~Kataoka, J.~Knödlseder, M.~Kuss, J.~Lande,
  S.~Larsson, L.~Latronico, F.~Longo, F.~Loparco, M.~N. Lovellette, P.~Lubrano,
  M.~N. Mazziotta, J.~E. McEnery, J.~Mehault, P.~F. Michelson, T.~Mizuno,
  C.~Monte, M.~E. Monzani, A.~Morselli, I.~V. Moskalenko, S.~Murgia,
  A.~Tramacere, E.~Nuss, J.~Greiner, M.~Ohno, T.~Ohsugi, N.~Omodei, M.~Orienti,
  E.~Orlando, J.~F. Ormes, D.~Paneque, J.~S. Perkins, M.~Pesce-Rollins, et~al.
\newblock The imprint of the extragalactic background light in the gamma-ray
  spectra of blazars.
\newblock {\em Science}, 338:1190--, 2012.

\bibitem{abr13}
H.E.S.S. Collaboration, A.~Abramowski, F.~Acero, F.~Aharonian, A.~G.
  Akhperjanian, G.~Anton, S.~Balenderan, A.~Balzer, A.~Barnacka, Y.~Becherini,
  J.~Becker~Tjus, K.~Bernlöhr, E.~Birsin, J.~Biteau, A.~Bochow, C.~Boisson,
  J.~Bolmont, P.~Bordas, J.~Brucker, F.~Brun, P.~Brun, T.~Bulik, S.~Carrigan,
  S.~Casanova, M.~Cerruti, P.~M. Chadwick, A.~Charbonnier, R.~C.~G. Chaves,
  A.~Cheesebrough, G.~Cologna, J.~Conrad, C.~Couturier, M.~Dalton, M.~K.
  Daniel, I.~D. Davids, B.~Degrange, C.~Deil, P.~deWilt, H.~J. Dickinson,
  A.~Djannati-Ataï, W.~Domainko, L.~O'C.~Drury, G.~Dubus, K.~Dutson, J.~Dyks,
  M.~Dyrda, K.~Egberts, P.~Eger, P.~Espigat, L.~Fallon, C.~Farnier, S.~Fegan,
  F.~Feinstein, M.~V. Fernandes, D.~Fernandez, A.~Fiasson, G.~Fontaine,
  A.~Förster, M.~Füßling, M.~Gajdus, Y.~A. Gallant, T.~Garrigoux, H.~Gast,
  B.~Giebels, J.~F. Glicenstein, B.~Glück, D.~Göring, M.-H. Grondin,
  S.~Häffner, J.~D. Hague, J.~Hahn, D.~Hampf, J.~Harris, S.~Heinz,
  G.~Heinzelmann, G.~Henri, G.~Hermann, A.~Hillert, J.~A. Hinton, W.~Hofmann,
  P.~Hofverberg, M.~Holler, D.~Horns, A.~Jacholkowska, C.~Jahn, M.~Jamrozy,
  I.~Jung, M.~A. Kastendieck, K.~Katarzyński, U.~Katz, S.~Kaufmann,
  B.~Khélifi, D.~Klochkov, W.~Kluźniak, T.~Kneiske, Nu. Komin, K.~Kosack,
  R.~Kossakowski, F.~Krayzel, H.~Laffon, et~al.
\newblock Measurement of the extragalactic background light imprint on the
  spectra of the brightest blazars observed with h.e.s.s.
\newblock {\em Astronomy and Astrophysics}, 550, 2013.

\bibitem{dea11}
Alessandro de~Angelis, Giorgio Galanti, and Marco Roncadelli.
\newblock Relevance of axionlike particles for very-high-energy astrophysics.
\newblock {\em Physical Review D}, 84:105030, 2011.

\bibitem{wou12}
Denis Wouters and Pierre Brun.
\newblock Irregularity in gamma ray source spectra as a signature of axionlike
  particles.
\newblock {\em Physical Review D}, 86:43005, 2012.

\bibitem{bru13}
Pierre Brun.
\newblock Axion-like particles: possible hints and constraints from the
  high-energy universe.
\newblock {\em Journal of Physics Conference Series}, 460:2015, 2013.

\bibitem{bit15}
J.~Biteau and D.~A. Williams.
\newblock The extragalactic background light, the hubble constant, and
  anomalies: Conclusions from 20 years of tev gamma-ray observations.
\newblock {\em The Astrophysical Journal}, 812, 2015.

\bibitem{sim08}
Melanie Simet, Dan Hooper, and Pasquale~D. Serpico.
\newblock Milky way as a kiloparsec-scale axionscope.
\newblock {\em Physical Review D}, 77:63001, 2008.

\bibitem{bur09}
Clare Burrage, Anne-Christine Davis, and Douglas~J. Shaw.
\newblock Active galactic nuclei shed light on axionlike particles.
\newblock {\em Physical Review Letters}, 102:201101, 2009.

\bibitem{mir09}
Alessandro Mirizzi and Daniele Montanino.
\newblock Stochastic conversions of tev photons into axion-like particles in
  extragalactic magnetic fields.
\newblock {\em Journal of Cosmology and Astro-Particle Physics}, 12:004, 2009.

\bibitem{fai11}
M.~Fairbairn, T.~Rashba, and S.~Troitsky.
\newblock Photon-axion mixing and ultra-high energy cosmic rays from bl lac
  type objects: Shining light through the universe.
\newblock {\em Physical Review D}, 84:125019, 2011.

\bibitem{hor12}
Dieter Horns, Luca Maccione, Manuel Meyer, Alessandro Mirizzi, Daniele
  Montanino, and Marco Roncadelli.
\newblock Hardening of tev gamma spectrum of active galactic nuclei in galaxy
  clusters by conversions of photons into axionlike particles.
\newblock {\em Physical Review D}, 86:75024, 2012.

\bibitem{abr13b}
A.~Abramowski, F.~Acero, F.~Aharonian, F.~Ait~Benkhali, A.~G. Akhperjanian,
  E.~Ang¨¹ner, G.~Anton, S.~Balenderan, A.~Balzer, A.~Barnacka, Y.~Becherini,
  J.~Becker~Tjus, K.~Bernl?hr, E.~Birsin, E.~Bissaldi, J.~Biteau, C.~Boisson,
  J.~Bolmont, P.~Bordas, J.~Brucker, F.~Brun, P.~Brun, T.~Bulik, S.~Carrigan,
  S.~Casanova, M.~Cerruti, P.~M. Chadwick, R.~Chalme-Calvet, R.~C.~G. Chaves,
  A.~Cheesebrough, M.~Chr¨¦tien, S.~Colafrancesco, G.~Cologna, J.~Conrad,
  C.~Couturier, M.~Dalton, M.~K. Daniel, I.~D. Davids, B.~Degrange, C.~Deil,
  P.~deWilt, H.~J. Dickinson, A.~Djannati-Ata?, W.~Domainko, L.~O.'C. Drury,
  G.~Dubus, K.~Dutson, J.~Dyks, M.~Dyrda, T.~Edwards, K.~Egberts, P.~Eger,
  P.~Espigat, C.~Farnier, S.~Fegan, F.~Feinstein, M.~V. Fernandes,
  D.~Fernandez, A.~Fiasson, G.~Fontaine, A.~F?rster, M.~F¨¹?ling, M.~Gajdus,
  Y.~A. Gallant, T.~Garrigoux, H.~Gast, B.~Giebels, J.~F. Glicenstein,
  D.~G?ring, M.-H. Grondin, M.~Grudzi¨½ska, S.~H?ffner, J.~D. Hague, J.~Hahn,
  J.~Harris, G.~Heinzelmann, G.~Henri, G.~Hermann, O.~Hervet, A.~Hillert, J.~A.
  Hinton, W.~Hofmann, P.~Hofverberg, M.~Holler, D.~Horns, A.~Jacholkowska,
  C.~Jahn, M.~Jamrozy, M.~Janiak, F.~Jankowsky, I.~Jung, M.~A. Kastendieck,
  K.~Katarzy¨½ski, U.~Katz, S.~Kaufmann, B.~Kh¨¦lifi, M.~Kieffer, S.~Klepser,
  D.~Klochkov, W.~Klu?niak, et~al.
\newblock Constraints on axionlike particles with h.e.s.s. from the
  irregularity of the pks 2155-304 energy spectrum.
\newblock {\em Physical Review D}, 88, 2013.

\bibitem{mey14}
Manuel Meyer, Daniele Montanino, and Jan Conrad.
\newblock On detecting oscillations of gamma rays into axion-like particles in
  turbulent and coherent magnetic fields.
\newblock {\em Journal of Cosmology and Astro-Particle Physics}, 09, 2014.

\bibitem{wou14}
Denis Wouters and Pierre Brun.
\newblock Anisotropy test of the axion-like particle universe opacity effect: a
  case for the cherenkov telescope array.
\newblock {\em Journal of Cosmology and Astro-Particle Physics}, 01, 2014.

\bibitem{gal15}
Giorgio Galanti, Marco Roncadelli, Alessandro De~Angelis, and Giovanni~F.
  Bignami.
\newblock Advantages of axion-like particles for the description of
  very-high-energy blazar spectra, March 1, 2015 2015.
\newblock 33 pages, 9 figures, additional material added.

\bibitem{bul14}
Esra Bulbul, Maxim Markevitch, Adam Foster, Randall~K. Smith, Michael
  Loewenstein, and Scott~W. Randall.
\newblock Detection of an unidentified emission line in the stacked x-ray
  spectrum of galaxy clusters.
\newblock {\em The Astrophysical Journal}, 789, 2014.

\bibitem{boy14}
A.~Boyarsky, O.~Ruchayskiy, D.~Iakubovskyi, and J.~Franse.
\newblock Unidentified line in x-ray spectra of the andromeda galaxy and
  perseus galaxy cluster.
\newblock {\em Physical Review Letters}, 113, 2014.

\bibitem{hig14}
Tetsutaro Higaki, Kwang~Sik Jeong, and Fuminobu Takahashi.
\newblock The 7 kev axion dark matter and the x-ray line signal.
\newblock {\em Physics Letters B}, 733:25--31, 2014.

\bibitem{jae14}
Joerg Jaeckel, Javier Redondo, and Andreas Ringwald.
\newblock 3.55Â kev hint for decaying axionlike particle dark matter.
\newblock {\em Physical Review D}, 89, 2014.

\bibitem{cic14}
Michele Cicoli, Joseph~P. Conlon, M.~C.~David Marsh, and Markus Rummel.
\newblock 3.55Â kev photon line and its morphology from a 3.55Â kev axionlike
  particle line.
\newblock {\em Physical Review D}, 90, 2014.

\bibitem{aba14}
Kevork~N. Abazajian.
\newblock Resonantly produced 7Â kev sterile neutrino dark matter models and
  the properties of milky way satellites.
\newblock {\em Physical Review Letters}, 112, 2014.

\bibitem{con14}
Joseph~P. Conlon and Francesca~V. Day.
\newblock 3.55 kev photon lines from axion to photon conversion in the milky
  way and m31.
\newblock {\em Journal of Cosmology and Astro-Particle Physics}, 11, 2014.

\bibitem{ade15}
P.~A.~R. {Ade}, K.~{Arnold}, M.~{Atlas}, C.~{Baccigalupi}, D.~{Barron},
  D.~{Boettger}, J.~{Borrill}, S.~{Chapman}, Y.~{Chinone}, A.~{Cukierman},
  M.~{Dobbs}, A.~{Ducout}, R.~{Dunner}, T.~{Elleflot}, J.~{Errard},
  G.~{Fabbian}, S.~{Feeney}, C.~{Feng}, A.~{Gilbert}, N.~{Goeckner-Wald},
  J.~{Groh}, G.~{Hall}, N.~W. {Halverson}, M.~{Hasegawa}, K.~{Hattori},
  M.~{Hazumi}, C.~{Hill}, W.~L. {Holzapfel}, Y.~{Hori}, L.~{Howe}, Y.~{Inoue},
  G.~C. {Jaehnig}, A.~H. {Jaffe}, O.~{Jeong}, N.~{Katayama}, J.~P. {Kaufman},
  B.~{Keating}, Z.~{Kermish}, R.~{Keskitalo}, T.~{Kisner}, A.~{Kusaka}, M.~{Le
  Jeune}, A.~T. {Lee}, E.~M. {Leitch}, D.~{Leon}, Y.~{Li}, E.~{Linder},
  L.~{Lowry}, F.~{Matsuda}, T.~{Matsumura}, N.~{Miller}, J.~{Montgomery}, M.~J.
  {Myers}, M.~{Navaroli}, H.~{Nishino}, T.~{Okamura}, H.~{Paar}, J.~{Peloton},
  L.~{Pogosian}, D.~{Poletti}, G.~{Puglisi}, C.~{Raum}, G.~{Rebeiz}, C.~L.
  {Reichardt}, P.~L. {Richards}, C.~{Ross}, K.~M. {Rotermund}, D.~E. {Schenck},
  B.~D. {Sherwin}, M.~{Shimon}, I.~{Shirley}, P.~{Siritanasak}, G.~{Smecher},
  N.~{Stebor}, B.~{Steinbach}, A.~{Suzuki}, J.-i. {Suzuki}, O.~{Tajima},
  S.~{Takakura}, A.~{Tikhomirov}, T.~{Tomaru}, N.~{Whitehorn}, B.~{Wilson},
  A.~{Yadav}, A.~{Zahn}, O.~{Zahn}, and {Polarbear Collaboration}.
\newblock {POLARBEAR constraints on cosmic birefringence and primordial
  magnetic fields}.
\newblock {\em Physical Review D}, 92(12):123509, December 2015.

\bibitem{gr01}
D.~{Grasso} and H.~R. {Rubinstein}.
\newblock {Magnetic fields in the early Universe}.
\newblock {\em Physics Reports}, 348:163--266, July 2001.

\bibitem{gro02}
Y.~Grossman, S.~Roy, and J.~Zupan.
\newblock Effects of initial axion production and photon-axion oscillation on
  type ia supernova dimming [rapid communication].
\newblock {\em Physics Letters B}, 543:23--28, 2002.

\bibitem{lai06}
Dong Lai and Jeremy Heyl.
\newblock Probing axions with radiation from magnetic stars.
\newblock {\em Physical Review D}, 74:123003, 2006.

\bibitem{kuo89}
T.~K. Kuo and James Pantaleone.
\newblock Neutrino oscillations in matter.
\newblock {\em Reviews of Modern Physics}, 61:937--980, 1989.

\end{thebibliography}

\end{document}